\documentclass[fleqn,usenatbib,useAMS]{mnras}

\usepackage{graphicx}	
\usepackage{amsmath}
\usepackage{amssymb}	
\usepackage{bm}
\usepackage{physics} 
\usepackage{multicol}
\usepackage{mathrsfs}
\usepackage{upgreek}

\newcommand{\p}{\partial}
\newcommand{\OmK}{\Omega_\text{K}}

\newcommand{\VA}{v_\mathrm{A}}

\newcommand{\bB}{\mathbf{B}}
\newcommand{\bI}{\mathbf{I}}
\newcommand{\bj}{\mathbf{j}}

\newcommand{\bE}{\mathbf{E}}

\newcommand{\bM}{\mathbf{M}}

\newcommand{\bv}{\mathbf{v}}

\newcommand{\bz}{\mathbf{z}}

\newcommand{\nn}{\nonumber}

\begin{document}
\title[]{Dynamics of cold circumstellar gas in debris disks}

\author[]{Can Cui$^{1}$\thanks{E-mail: \href{mailto:cc795@cam.ac.uk}{can.cui@astro.utoronto.ca}}, Sebastian Marino$^{2}$, Quentin Kral$^{3}$ and Henrik Latter$^{1}$   \\ \\
$^{1}$DAMTP, University of Cambridge, Wilberforce Road, Cambridge CB3 0WA, UK \\
$^{2}$Department of Physics and Astronomy, University of Exeter, Stocker Road, Exeter EX4 4QL, UK \\
$^{3}$LESIA, Observatoire de Paris, Université PSL, CNRS, Sorbonne Université, Université Paris Cité, 5 place Jules Janssen, \\ 92195
Meudon, France
}

\pubyear{2023}

\label{firstpage}
\pagerange{\pageref{firstpage}--\pageref{lastpage}}
\maketitle

\begin{abstract}

Mounting observational evidence indicates that cold circumstellar gas is present in debris disk systems. This work focuses on various dynamical processes that debris-disk gas may undergo. We review five mechanisms that can transport angular momentum and their applications to debris disks. These include molecular viscosity, hydrodynamic turbulence, magnetohydrodynamic turbulence, magnetized disk winds, and laminar magnetic stress. 
We find that molecular viscosity can result in $\alpha$ as high as $\lesssim 0.1$ for sufficiently low densities, while the Rossby wave instability is a possible source of hydrodynamic turbulence and structure formation. We argue that the vertical shear instability is unlikely due to the long cooling times. The onset of the magnetorotational instability (MRI) is dichotomous: for low density disks the MRI can be excited at the midplane, while for high mass disks it may only be operating at $z>2-3H$, if at all. The MHD wind and laminar magnetic stress mechanisms rely on the configuration and strength of any background large-scale magnetic field, the existence of which is uncertain and possibly unlikely. 
We conclude that the dominant mechanism and its efficiency in transporting angular momentum varies from one system to the other, depending especially closely on the gas density. More detailed analyses shall be performed in the future focusing on representative, nearby debris disks.   

\end{abstract}


\section{Introduction}\label{in}

Main sequence stars are commonly orbited by solid circumstellar material, such as planets and debris \citep{wyatt+08,matthews+14,hughes+18}. The latter is made up of planetesimals as well as the dust and gas derived from them. Debris disks are tenuous, optically-thin (with respect to dust), and may persist over giga years. The nearest debris disk is the Solar System's asteroid and Kuiper belts. Extrasolar systems also commonly harbor debris disks: to name a few, Fomalhaut, $\beta$ Pictoris, Vega, and $\epsilon$ Eridani \citep[e.g.,][]{holland+98}. 

Originally, the absence of gas was a distinguishing characteristic of debris disks. However, mounting observational evidence in recent years indicates that cold circumstellar gas is present around a large fraction of stars hosting debris disks, identified mainly via CO gas emission at millimeter wavelengths \citep[see][for recent reviews]{strom+20, marino22}. Large quantities of CO gas were found mainly around young A-type stars initially \citep[e.g.][]{zuckerman+12, kospal+13,dent+14, lieman-sifry+16, moor+17}, while later and deeper observations detected CO gas at lower levels around later spectral types \citep{marino+16, matra+19twa7, kral+20b}. CO is known to be short-lived due to photodissociation driven by the UV interstellar radiation field, unless shielded, and its dissociation products have also been found in some of these systems: CI, CII, OI \citep[e.g.][]{roberge+06, cataldi+14,Cataldi+18,kral+19}. ALMA observations of these species have shown that the gas and dust spatial distributions overlay, with large inner cavities of tens of AU in radius \citep[e.g.][]{kospal+13, dent+14, marino+16, moor+17}. Some of the observations also suggest that the gas is radially spreading relative to the dust \citep[e.g.,][]{kospal+13,moor+13}. 

The origin of the circumstellar gas still remains unclear, particularly in very massive gas disks. Both primordial and secondary origin scenarios have been proposed to explain some of the observed features. The primordial origin idea proposes that the circumstellar gas is inherited from earlier Class II protoplanetary disks, and hence is expected to be H$_2$-dominated \citep[e.g.,][]{moor+17,nakatani_etal20,Klusmeyer+21,Smirnov-Pinchukov+22}. Meanwhile, the secondary origin idea suggests that the gas is released by solid bodies in the planetesimal belt, via the grinding down of larger volatile-rich solid bodies in a collisional cascade, sublimation by external heating, or radionuclides; the gas is hence H$_2$-poor \citep[e.g.,][]{Czechowski+07,moor+11,zuckerman+12,dent+14,matra+17}.

Though tenuous, circumstellar gas has been invoked in a number of physical processes in debris disk systems. Gas can influence dust dynamics and shape its spatial distribution \citep[e.g.,][]{takeuchi01, krivov+09, olofsson+22}. The photoelectric instability, whose onset depends on gas radial pressure gradients, can concentrate dust into annular structures when the disk is optically thin to starlight \citep{lyra+13}. Circumstellar gas can also accrete onto neighboring planets and hence contribute to the chemical composition of secondary planetary atmospheres \citep{kral+20}. Finally, it has been proposed that gas dynamics might be the origin of the dust (and gas) clump detected in $\beta$ Pic, via a gas vortex triggered by the Rossby wave instability \citep{Skaf+23}. 

Despite its importance in many physical processes, the understanding of gas dynamics in debris disks is relatively under-developed \cite[but see][for low gas density disk of $\beta$ Pic]{kl16}. Perhaps the most salient dynamical questions in this context concern the radial spreading and vertical mixing of gas. The former is closely related to angular momentum transport, and the latter strongly influences the degree of midplane shielding of CO by surface CI, which in turn helps determine the origin of the gas. To this aim, this paper seeks solutions to some fundamental questions regarding gas dynamics. These include: what are the nature and strength of the angular momentum transport? Are hydrodynamic mechanisms important, such as molecular viscosity and hydrodynamic turbulence? If the gas dynamics is controlled by magnetohydrodynamics (MHD), is the disk MRI turbulent or do magnetized disk winds dominate mass accretion? Does non-ideal MHD effects come into play? In short, what is the origin of any `anomalous viscosity' in debris disks, and can a simple Shakura-Sunyaev $\alpha$-parameter describe the underlying physical processes adequately? 

The paper is organized as follows. In \S\ref{sec:bg}, we start with a background introduction on gas observations, theoretical modeling, and a comparison of gas physical properties between protoplanetary and debris disks. In \S\ref{sec:marino22}, our methods are outlined. In \S\ref{sec:hydro}, we introduce hydrodynamic mechanisms that may shape gas structures. In \S\ref{sec:MHD}, we investigate the MHD effects on the gas dynamics. Finally, we discuss the main results in \S\ref{sec:disc}.

\section{Background}\label{sec:bg}

\begin{table*}
\caption{Comparison of physical parameters between protoplanetary and debris disks at 100 AU (in cgs units).} 
\begin{tabular*}{1\textwidth}{l@{\hspace{3cm}}l@{\hspace{3cm}}l@{\hspace{3cm}}l@{\hspace{3cm}}}
\hline
\hline
parameter & symbol & protoplanetary & debris  \\
\hline
gas number density & $n$                & $10^5-10^9$  & $10^2-10^6$  \\
mean free path     & $l/H$              & $10^{-8}-10^{-4}$         & $10^{-5}-1$ \\
ionization         & $x_\mathrm{e}$     & $10^{-13}-10^{-10}$ &  $10^{-7}-1$    \\   
dust-to-gas ratio  & $\rho_\mathrm{d}/\rho_\mathrm{g}$     & $10^{-2}$  & $1-10^5$   \\
Stokes number      & St                 & $< 1$            & $>1$   \\ 
\hline
\end{tabular*}
\label{table:com} 
\end{table*}

\subsection{Observations of cold gas}\label{sec:obs}

The main observed species at tens of AU have been CO, CI, CII, and OI. CO has a longer photodissociation lifetime compared to other molecules, and its bright rotational lines at millimeter wavelengths are where ALMA is most sensitive. CI most likely originates from the photodissociation of CO and is long-lived, allowing resolved observations with ALMA. Additionally, CII and OI have been detected by the Herschel far-IR survey in a few systems. Of these observable species, the gas mass tends to be dominated by CO, except that $\beta$~Pic and 49~Ceti may have slightly higher CI masses \citep{cataldi+23}. 

There is a wide distribution of CO masses from ${\sim0.1}\ M_{\oplus}$ down to marginally detected levels of $10^{-7}\ M_{\oplus}$ \citep[e.g.][]{matra+17b}. The high end of the distribution, $M_{\rm CO} \sim 10^{-1}-10^{-4}\ M_{\oplus}$, is dominated by A-type stars with bright debris discs \citep{marino+20}, while the lower end of the distribution contains systems with a wide range of spectral types, from B- to M-type stars \citep[e.g.][]{marino+16, kral+20, matra+19twa7, rebollido+22}. Age and dust levels also play a factor. Massive gas discs with $M_{\rm CO} > 10^{-4}\ M_{\oplus}$ are found mostly around A-type stars younger than 50~Myr, and with dust fractional luminosities above $10^{-4}$ \citep{moor+17}. However, this might be due to observational biases since no surveys have been dedicated to searching for CO around older and less dusty discs. Among the surveyed dust-rich debris disks around young A-type stars, CO is readily detected in 11/17; 10 of which have CO gas masses above $10^{-4}\ M_{\oplus}$ \citep{moor+17, kennedy+18, hales+22}. 

Despite the abundance of gas, its bulk composition is still largely unknown. In particular, there are no strong constraints on the presence of H or H$_2$ in the most massive systems, which would dominate the mass if the gas was of primordial origin. For lower mass systems, it is thought that the H$_2$ content is low because of the subthermal gas excitation \citep{matra+17}. There have been multiple attempts to detect carbon- and oxygen-bearing molecules other than CO that are frequently detected in protoplanetary disks, such as CN, HCN, HCO$^{+}$, CCH, CH$_{3}$OH, CS, and SiO \citep[][]{Klusmeyer+21, Smirnov-Pinchukov+22}. However, none have been found, ruling out abundances relative to CO similar to protoplanetary disks. These low abundances, however, can be simply explained by the optically thin nature of debris disks: the UV stellar and interstellar radiation can penetrate the disk and quickly dissociate most molecules, even if the gas is ISM-like with a high abundance of H or H$_{2}$. 

\subsection{1D modeling of secondary gas evolution}\label{sec:modeling}

Previously modeling of the gas evolution has focused on two issues: how gas viscously expands radially, forming an accretion disc, and how CO photodissociates, producing atomic carbon and oxygen. 

To model the radial viscous expansion, two free parameters, the Shakura-Sunyaev $\alpha$ and the CO gas production rate, have been generically employed and varied,  to reproduce CO gas surface densities and radial distributions.
\cite{kral+16, kral+17} presented a 1D radial viscous evolution model to predict the CO, CI, CII and OI gas distribution and abundances. This model simulated how gas released at a fixed rate at the planetesimal belt location will viscously spread, using an $\alpha$ model, ionize and heat or cool. 
Subsequently, it was realized that CO could be shielded by CI, and thus if enough CI accumulates, the shielding could protect CO and explain the rich CO detected around many A-type stars \citep{kral+19}. This realization meant that all gas found in debris disks could be of secondary origin. \citet{marino+20} incorporated this into a 1D radial evolution model to produce synthetic populations of discs where the gas production decreases with age as discs lose mass. Comparing these against observations found that the distribution of CO masses is best explained by $\alpha=0.1$. 

Besides the viscous expansion in radius, the gas can be blown out over short timescales and persist beyond the planetesimal belt by stellar wind protons, creating belt winds of CO, CO$^+$, CI and OI \citep{kral+21,kral+23}. This occurs when the gas density is below a critical threshold, for example, in the least massive gas disks detected so far.

More recently, attention has been drawn to the vertical distribution of the gas \citep{olofsson+22}. The gas' precise vertical structure significantly influences the lifetime of CO, and thus on how much CO gas a disc can accumulate. The vertical mixing is determined by the vertical turbulent diffusion. If the gas is well mixed, \cite{cataldi+20} showed that CI shielding is weak and the CO lifetime increases only linearly with the column density of CI. Instead, if the gas is not mixed, where CI is distributed on a layer high above the midplane and above the CO gas, it can shield CO more effectively thus increasing the lifetime of CO exponentially \citep{kral+19}.

\citet{marino+22} investigated the vertical distribution of gas in the secondary origin scenario, using a 1D model that resolves the gas disc vertically. They showed that both scenarios are possible depending on how strong vertical mixing is, parametrized by a vertical (dimensionless) diffusivity $\alpha_v$. If the vertical mixing is efficient, CO and CI would have similar vertical distributions and the shielding by CI would weaken. If the vertical diffusivity is much weaker than the radial, and gas is released at a high rate, then a layered structure can be formed. This is a consequence of the disc becoming optically thick to photodissociating UV photons and the vertical diffusion timescale being shorter than the timescale at which gas is removed or replenished. In this case, the CO dominates the midplane, the CI density peaks above the midplane, at a height that depends on the gas surface density, and a low-density layer of CII dominates the uppermost layers.

\subsection{Protoplanetary vs debris disks}\label{sec:com}

Most previous work, as discussed in the last subsection, explicitly includes the effects of dynamical processes via simple parametrisations (such as by dimensionless diffusivities, $\alpha$, $\alpha_v$), but without specifying the nature of those mechanisms or interrogating the accuracy of the parametrisations. It is the goal of this paper to explore the poorly understood dynamics of debris-disk gas in more detail. To make inroads here, it is almost unavoidable to first consider protoplanetary disks and their rich dynamical behaviour. To help facilitate this, it is beneficial to compare the key characteristic physical and dynamical parameters of the two systems. These parameters and respective values are listed in Table \ref{table:com}. To make easy comparison, the values are calculated at 100 AU and vary from the midplane to the surface.

Protoplanetary disks are gas-rich and mainly composed of H$_2$. The gas surface density is about $10^2$~g~cm$^{-2}$ and number density of $10^5-10^9$~cm$^{-3}$ at 100 AU \citep{weidenschilling77,henning13}. Debris disks are much more dilute. The surface density of CO typically peaks at 100 AU and varies from $10^{-7}-10^{-6}$~g~cm$^{-2}$ for tenuous disks such as $\beta$ Pic, to $10^{-6}-10^{-4}$~g~cm$^{-2}$ for dense disks such as 49 Ceti \citep{kl16,hughes+17,higuchi+20,cataldi+23}.
The gas number density is about $10^2-10^7$ cm$^{-3}$ (\S\ref{sec:if}). Note that the high end of the range of CO surface densities could be much higher. This is because  $^{12}$CO emission is optically thick and CO isotopologue-selective photodissociation is usually ignored when estimating CO gas masses. The low end, on the other hand, is determined by the current sensitivity limits.

The gas mean free path determines the collisionality and molecular viscosity. Protoplanetary disks are generically neutral and collisional. Taking the cross section of H$_2$ to be approximately $10^{-15}$~cm$^2$, the mean free path is $l/H\sim 10^{-8}-10^{-4}$, where $l$ is the gas mean free path and $H$ the pressure scale height. Debris disks possess a non-negligible fraction of ionized gas (CII). The mean free path for CII-CII collisions is short due to the Coulomb encounters, and $l/H\sim 10^{-8}-10^{-4}$. On the other hand, neutral-neutral and neutral-ion collisions have longer mean free paths, resulting in $l/H\sim 10^{-5}-1$ \citep[][\S\ref{sec:mw}]{kral+16}. 
The ionization fraction is a key parameter for MHD processes, defined as the ratio of electron number density to neutral number density, $x_\mathrm{e}=n_\mathrm{e}/n$.
The ionization fraction is extremely low for protoplanetary disks, $x_\mathrm{e}\sim10^{-13}-10^{-10}$ \citep{lesur21}, whereas it is substantially higher in debris disks, $x_\mathrm{e}\sim10^{-7}-1$ \citep[Figure \ref{fig:seba};][]{kral+16, marino22}. Rough observational constraints have been placed on the ionization fraction in some debris disk systems and found to be $x_\mathrm{e}\sim 0.1-1$ based on CI and CII \citep{cataldi+23}.

Aerodynamic coupling between dust and gas is subject to a mutual drag force. The force is determined by the dust-to-gas density ratio and the dimensionless Stokes number. Protoplanetary disks are gas-dominated; the dust-to-gas ratio reaches only $1\%$. On the other hand, debris disks are dust-dominated; the dust-to-gas (CO) ratio can vary from unity to $10^5$ \citep{rebollido+22}, though this quantity is uncertain because it depends on modelling assumptions, such as the size of the largest solids. The Stokes number, St, is defined as the ratio of the stopping time to the dynamical time. The stopping time is the time scale required for the dust to respond to differences in the gas's and dust's velocities; the dynamical time is the local orbital period of the gas. When St is smaller (greater) than unity, the dust and gas are dynamically well (poorly) coupled. For micron- to millimeter-sized dust at the midplane and at $100$ AU in protoplanetary disks, $\mathrm{St}<1$. For the same sized dust in debris disks, generally $\mathrm{St}>1$, because of their considerably lower gas densities \citep{marino+20,huang23}.  

\subsection{Angular momentum transport theory}\label{sec:amt}

To better understand angular momentum transport in debris disks, and thus how they spread and accrete, we briefly introduce its underlying mathematical theory.
The momentum equation in conservative form is \citep{landau59}
\begin{equation}
\frac{\partial(\rho\bv)}{\partial t} + \nabla\cdot\bM = 0,
\end{equation}
where the momentum flux tensor is defined as
\begin{equation}
\vb{M} \equiv \rho\vb{v}\vb{v}-\frac{\vb{B}\vb{B}}{4\pi}+\bigg(P+\frac{B^2}{8\pi}\bigg)\bI - \btau,
\label{eq:tensor}
\end{equation}  
in which $\bv$ and $\bB$ are the gas velocity and magnetic field vector, $\bI$ is the identity tensor, and $\btau$ is the viscous stress tensor,
\begin{equation}
\btau=\eta[\nabla\bv+(\nabla\bv)^T]-\frac{2}{3}\eta(\nabla\cdot\bv)\bI .
\end{equation} 
The dynamical viscosity is denoted by $\eta=\rho\nu$, and $\nu$ is the kinematic viscosity. In cylindrical coordinates, the viscous stress tensor is $\tau_{R\phi}=\eta R\p\Omega/\p R$ for a Keplerian disk with angular speed $\Omega$. Note that we have omitted the dust drag on the gas in this model, and throughout the paper. This can be justified for a large class of debris disks \citep[see][]{kl16,kral+16}, but when the dust-to-gas ratio is large the drag term must be reinstated.

Under axisymmetry, the $R\phi$ and $z\phi$ components of the total stress tensor $\mathbf{T}$ are
\begin{equation}
T_{R\phi} = \rho\delta v_R\delta v_\phi - \frac{B_RB_\phi}{4\pi} + \frac{3}{2}\eta\Omega, 
\label{eq:rp}
\end{equation}
\begin{equation}
T_{z\phi}=  \rho v_zv_\phi- \frac{B_z B_\phi}{4\pi},
\label{eq:zp}
\end{equation}
where $\delta v_R$, $\delta v_\phi$ are the radial and azimuthal velocity fluctuations, respectively. 

From above, the angular momentum can be transported radially and vertically \citep[see reviews by][]{armitage11,wbf21,lesur+23}. The $T_{R\phi}$ term facilitates the radial transport of angular momentum, where part of the gas loses angular momentum and accretes onto the central star, and the rest gains angular momentum and expands radially outwards. 
The $\alpha$-disk model introduces a dimensionless parameter to quantify the strength of the $R\phi$ stress, 
\begin{equation}
\alpha = \frac{{\overline{T_R\phi}}}{\overline{P}}, 
\label{eq:alpha}
\end{equation}
where $P$ is the gas pressure and an overbar denotes a vertical average. The $T_{z\phi}$ term represents the vertical transport of angular momentum. Magnetized disk winds can drive such vertical transport. They extract angular momentum from the disk by exerting a torque at the disk surface, forcing the entire disk to accrete. 

Five mechanisms can be identified that contribute to the angular momentum transport process: molecular viscosity, hydrodynamic turbulence, magnetic turbulence, magnetized disk winds, and laminar magnetic stress. The molecular viscosity can transfer angular momentum through the last term of the $T_{R\phi}$ stress. Turbulence contributes to transport via the $T_{R\phi}$ stress. Hydrodynamic turbulence only manifests in the first term of $T_{R\phi}$. One example might be that arises from the vertical shear instability (\S\ref{sec:vsi}). Magnetic turbulence can produce both the first and second terms in $T_{R\phi}$. One well known example is the magneto-rotational instability (\S\ref{sec:amt_mhd}). On the other hand, a large-scale laminar magnetic stress can contribute to the angular momentum transport through the second term in $T_{R\phi}$. The second term of $T_{z\phi}$ arises from large-scale poloidal magnetic fields, which may permit the launching of magnetized disk winds. In the following sections we will assess the viability of each of these mechanisms in debris disk gas.

\section{Methods} \label{sec:marino22}

To help us investigate the different hydrodynamic and MHD processes going on in debris-disk gas, we require a disk model describing the spatial distribution of the gas species. Our reference disks in this paper have been calculated by \citet{marino+22}, the details of which we now describe. 

The model of \citet{marino+22} follows the vertical evolution of gas using a 1D grid that resolves the vertical axis. It considers a belt centred at 100~AU orbiting a 1.5~M$_\odot$ star. The model accounts for the release of CO gas from a planetesimal belt at different release rates, and it then follows how photodissociation and ionization by interstellar UV and radial viscous spreading and vertical mixing shape the vertical distribution of gas species. The viscous evolution and vertical turbulent mixing are parameterized by the Shakura-Sunyaev $\alpha$ parameter. The model adopted in our paper employs radial and vertical diffusion coefficients of $\alpha=10^{-2}$ and $\alpha_v=10^{-2}$, respectively.
The disk aspect ratio is set to $H/R=0.05$, equivalent to a temperature of $\sim50$ K. 
The mean molecular weight $\mu$ in their calculation is assumed to be equal to 14, equivalent to a gas dominated by carbon and oxygen in equal proportions.

A key parameter in the model is the solids' gas release rate, which varies between $10^{-3}$ and $10^1$ M$_\oplus$/Myr. This is a wide range that covers the expected release rate in known systems with detected CO gas, from the bottom of the distribution \citep[${\sim}10^{-2}$~M$_\oplus$/Myr, e.g. HD~181327,][]{marino+16} to the most massive gas-rich systems \citep[${\sim}1$~M$_\oplus$/Myr, e.g. HD~121617][]{kral+19}. Rates as high as 10~M$_\oplus$/Myr might be plausible for a few Myr depending on the mass of the belt and the processes that drive the release of CO from solids \citep{bonsor+23}. 
These rates map straightforwardly onto the steady state total gas surface densities at 100 AU, namely  $\Sigma\sim 1.7\times10^{-6}-1.7\times10^{-2}$~g~cm$^{-2}$, once $\alpha$ is specified. Note that the high end of this range is a hundred times larger than the highest CO surface densities observed using ALMA ($\sim 2\times10^{-4}$~g~cm$^{-2}$) \citep{cataldi+23}. However, those observational estimates only account for CO and ignore the possibility of CO isotopologue-selective photodissociation. This means that the gas surface densities could be orders of magnitude higher. 

We note that \citet{marino+22} treats the $\alpha$ and $\alpha_v$ as free parameters that are independent of mass production rate, density, temperature, radius, height, etc. However, as we will show in \S\ref{sec:mw}, \S\ref{sec:vsi}, and \S\ref{sec:amt_mhd}, molecular viscosity, the vertical shear instability, and the magneto-rotational instability all result in different values of $\alpha$ over density and space. This breaking of the connection between the dynamical parameters and the disk thermodynamic state may lead to inconsistencies and possibly obscure important trends in the modelling. Future global numerical simulations may be needed to self-consistently and thus reliably obtain the values of $\alpha$.

Finally, \citet{marino+22} investigates various vertical diffusion coefficients $\alpha_v$, ranging from $10^{-5}$ to $10^{-2}$, to explore the vertical distributions of gas species as $\alpha_v/\alpha$ varies. While our chosen value for $\alpha$  and $\alpha_v$  of $10^{-2}$ may seem high and arbitrary, we note that our key results are not sensitive to this choice but rather to the surface density of gas (set by the gas release rate and $\alpha$ ).  A weaker $\alpha_v$ relative to $\alpha$ does not quantitatively change our main conclusions.


\section{Hydrodynamics}\label{sec:hydro}

In this section we present hydrodynamic mechanisms of angular momentum transport that may be relevant in the context of debris disks. Molecular viscosity, which can be of consequence due to the low gas densities, is discussed first, in \S\ref{sec:mw}. Two hydrodynamic instabilities that have the potential to prevail in debris disks are also explored: the vertical shear instability (VSI; \S\ref{sec:vsi}) and the Rossby wave instability (RWI; \S\ref{sec:rwi}).

\subsection{Non-negligible molecular transport}\label{sec:mw}

As described in \S\ref{sec:amt}, angular momentum can be transported by molecular viscosity, manifesting in the last term of eq. \eqref{eq:rp}. The kinematic viscosity can be estimated as $\nu\sim l c_s$, where $l$ and $c_s$ are the mean free path and sound speed of the molecules, respectively. Hence, $\alpha\sim \Omega\rho\nu/P\sim l/H$. Transport by molecular viscosity requires some degree of collisionality, and the estimate for $\nu$ above holds only if particle trajectories between collisions are not overly influenced by body forces and thus follow straight lines. When $l<H$, these restrictions are satisfied. But when $l>H$, gas particles will tend to undergo vertical and radial epicyclic oscillation with a characteristic length-scale $\sim H$, and a different estimate for $\nu$ is needed \citep[see][]{GP78,pringle81}.

We first compute the mean free path for different gas species, including collisions among neutrals (CO, CI, OI) and ions (CII). The collisions with electrons are ignored due to their negligible inertia compared to other gas species. The mean free path is $l=1/(n\sigma)$, where $n$ is the number density of the impacted species, and $\sigma$ is the relevant cross section of the interaction. 
A rigorous calculation of the mean free path requires the cross section for each projectile-target pair. Here, we simplify the calculation by considering three cases: neutral-neutral collision, neutral-CII collision, and CII-CII collision. For CII-CII collisions, $\sigma=\pi b^2$ is taken as the Coulomb cross section, where $b= e^2/(4\pi\epsilon_0 m_\mu c_s^2)$, $m_\mu$ the reduced mass of the two colliders, and $c_s$ the sound speed. For neutral-neutral and neutral-CII collisions, the cross section is approximated as geometric, $\sigma\approx\pi$ (10$^{-8}$ cm)$^2$. As the Coulomb impact parameter $b$ is orders of magnitude larger than the atomic size of the gas species, the mean free path of CII-CII collisions is much shorter than the other two. Thus, neutral-neutral and neutral-CII collisions result in more efficient angular momentum transport than CII-CII collisions, noting also that the neutral species never comprise less than half the gas mass.  

\begin{figure}
    \centering
    \includegraphics[width=0.5\textwidth]{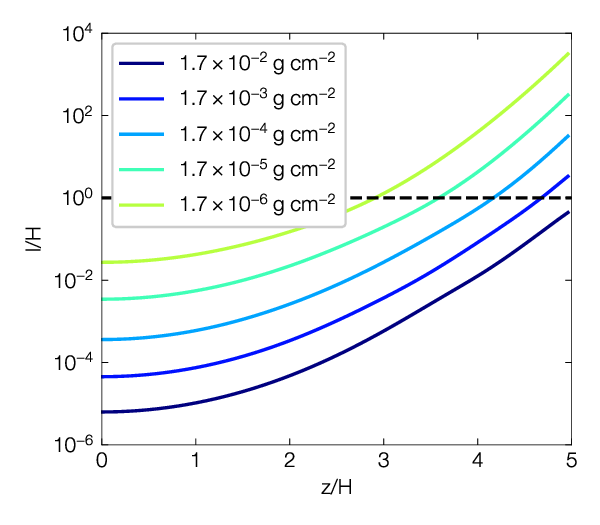}
    \caption{Neutral gas mean free path $l$ divided by pressure scale height $H$ as a function of height. Colors represent different gas surface densities. The horizontal dashed line denotes $l/H=1$.}
    \label{fig:coll}    
\end{figure}

In Figure \ref{fig:coll}, we plot the neutral mean free path $l$ divided by the gas pressure scale height $H$ as a function of height. The pressure scale height is calculated by assuming a constant temperature $T=50$ K with height throughout the paper. As is clear, in the bulk of the gas disk the mean free path is much less than the scale height and the assumptions made above are validated. Even the most dilute disks in our model sample are collisional towards the midplane, in agreement with previous work \citep{kral+16}. Note that values of $l\gtrsim H$ (above the dashed horizontal line) are inaccurate because of the dominance of body forces in this weakly collisional regime. 

Next, by vertically averaging the dynamical viscosity for $l<H$, we obtain the molecular viscous alpha by equation \eqref{eq:alpha}. We find $\alpha\sim 3.2\times10^{-5}$, $2\times10^{-4}$, $1.6\times10^{-3}$, $1.2\times10^{-2}$, and $9\times10^{-2}$ for the five surface densities selected in Figure \ref{fig:coll}, ranging from high to low, respectively. Lower density results in longer mean free paths, and thus enhanced transport. 
But note, as mentioned above, in more dilute disks, which are weakly collisional at all heights, gas particles undergo confined epicyclic oscillation before colliding and thus molecular transport, and the viscosity, will be lower \citep[section 4;][]{pringle81}.

The molecular viscosity sets a non-negligible lower limit for angular momentum transport in debris-disk gas. If the disk is completely laminar, exhibiting no other hydrodynamic or MHD process, then the disk will still spread and evolve under the action of its intrinsic molecular viscosity. Note that in such a situation the disk may be viscously unstable, due to the inverse proportionality between $\nu$ and surface density, and may possibly break up into rings and gaps \citep{schmidt09}. The timescale for this process is of order $(\alpha\Omega)^{-1}$, and thus might be relevant in more dilute systems.
Note also that this background low level of molecular viscosity could interfere with other mechanisms that can drive angular momentum transport, for example, the onset and development of the hydrodynamic instabilities described later. However, we exclude its impact in this work for simplicity. 

Finally, molecular transport may also give rise to significant vertical mixing, and thus to a non-turbulent $\alpha_v$. Naive estimates from kinetic theory would suggest $\alpha_v$ is of the same order of magnitude as the molecular $\alpha$. As a consequence, in more dilute system it could provide a non-trivial contribution to the vertical transport, in line with our adoption of the fiducial estimate $\alpha_v=10^{-2}$.

\subsection{Vertical shear instability}\label{sec:vsi}

The VSI is a promising hydrodynamic instability that may be operating in debris disk gas \citep{nelson_etal13,lp18,cl21,cl22,lk22}. Two conditions must be met for the onset of the VSI: a vertical gradient in the angular velocity, i.e. $d\Omega/dz\neq 0$, and a sufficiently short thermal relaxation timescale $t_c$. The former is readily satisfied in a thin locally isothermal disk if there is a radial temperature gradient \citep{nelson_etal13}. Numerical simulations of the VSI conducted, in the protoplanetary disk context, yield a Shakura-Sunyaev $\alpha$ ranging from $\sim10^{-3}$ with locally isothermal equation of state \citep{nelson_etal13}, down to $\sim10^{-5}-10^{-4}$ when including realistic radiative transport \citep{sk16,mario+17}. Non-ideal MHD simulations of VSI, applicable to the outer regions of protoplanetary disks, suggest $\alpha\sim10^{-4}-10^{-3}$ \citep{cb20,cb22}.

The critical thermal relaxation timescale for the onset of the VSI was derived by \citet{ly15} in a vertically global and radially local disk model. They show that the instability criterion can be framed as
\begin{equation}
t_c < \frac{|q|h}{(\gamma-1)\OmK} \ , 
\end{equation}
where $h=H/R$ is the disk aspect ratio, $q$ is the radial power-law index of the temperature, and $\gamma$ is the adiabatic index. In the context of debris disks, using $h\approx0.05$ for $T=50$ K at 100 AU, $q\approx -0.5$, and $\gamma\approx 1.4$, we find $t_c\lesssim0.01P_\mathrm{orb}$, where $P_\mathrm{orb}=2\pi/\OmK$ is the local orbital time scale.

To estimate $t_c$ in debris disk gas, we assume that the thermal relaxation rate in the disk is of order the heating rate. To estimate the latter, we assume that heating is dominated by the dust's photoelectric effect \citep{zagorovsky+10}. This assumption will produce an upper bound on the thermal relaxation timescale because, if enough carbon gas is in the system, photoionization heating may take over \citep{kral+16}. We adopt the photoelectric heating rate of eq. (23) in \citet{zagorovsky+10}, and divide it by the total internal energy of the gas, $\int e dV \sim \int\rho(z)\mathbb{R}T_\mathrm{gas}dV$, where $\mathbb{R}$ is the gas constant.
The parameters taken are: 
$T_\mathrm{gas}=50$~K, $R=100$~AU, $L_\mathrm{dust}/L_\mathrm{\ast}=10^{-3}$, $e\phi=1$~eV, and the number density of electrons are from \citet{marino+22}'s model. We leave to future work a more accurate and detailed calculation of the disk thermal physics.
 
\begin{figure}
    \centering
    \includegraphics[width=0.5\textwidth]{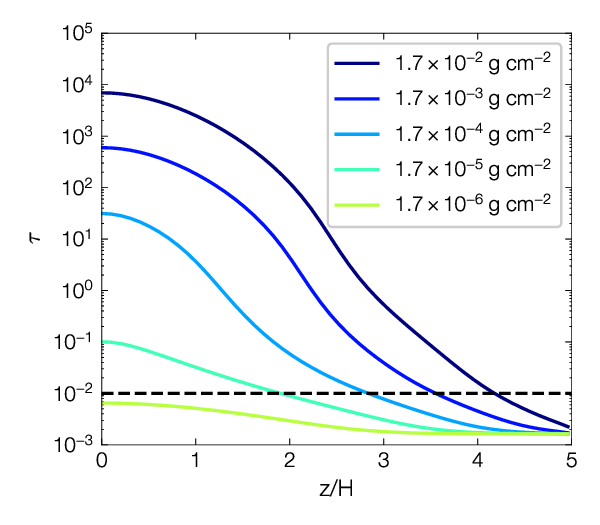}
    \caption{Thermal relaxation timescale $\tau=t_\mathrm{c}/P_\mathrm{orb}$ as a function of height, assuming cooling rate is balanced by photoelectric heating rate of dust grains. Dashed line is the VSI critical thermal relaxation timescale for $h=0.05$, $q=-0.5$, and $\gamma=1.4$ at $100$~AU.}
    \label{fig:cool}    
\end{figure}

 Figure \ref{fig:cool} shows the dimensionless thermal relaxation timescale $\tau=t_c/P_\text{orb}$. In the bulk of the disk $|z|< H$, the timescale is too long for the VSI to develop except for the most dilute disk in our sample. The densest disks possess midplane relaxation times many orders of magnitudes greater than the dynamical timescale. In all cases, at sufficiently high altitudes the relaxation time reaches a minimum value of $\tau\sim 0.01$, where the VSI may be excited, though if it does appear it will be in a form different to that appearing in protoplanetary disks, and unlikely to transport angular momentum effectively.

Finally, the above discussion is based on pure hydrodynamics, but magnetic fields will negatively impact on the onset of the VSI directly, via magnetic tension, and indirectly, via competing MRI turbulence \citep{lk22}. The MRI will be discussed in \S\ref{sec:MHD}. We also omit the effect of dust drag, which may be important in disks with large dust-to-gas ratios \citep{kl16} and could provide a strong stabilising effect \citep[cf.][]{ly17,LL21}.

\subsection{Rossby wave instability}\label{sec:rwi}

The 1D modeling of radial gas spreading in debris disks typically exhibits a bump structure in the surface density. This provides a natural precondition for the Rossby wave instability (RWI). The RWI \citep{lovelace99} is considered one way to generate large-scale vortices, and associated spiral density waves, in the context of accretion disks \citep{lovelace+14}. 
In barotropic disks, a necessary condition for instability is an extremum in the vortensity profile \citep{lovelace99,li_etal00}
\begin{equation}
\zeta=\frac{\hat{\bz}\cdot(\nabla\times\bv)}{\Sigma_g}=\frac{1}{\Sigma_g R}\frac{\p (R^2\Omega)}{\p R},
\label{eq:F}
\end{equation}
where $\Sigma_g=\int\rho\, dz$ is the surface density. 

One prominent feature of the RWI's non-linear evolution is the formation of anticyclonic vortices near the initial surface density extremum \citep{li+01}. Depending on the most unstable azimuthal modes, multiple vortices will emerge but eventually merge into one single giant vortex \citep{ono_etal16,ono+18}. These vortices enclose a pressure maximum which can trap dust grains in azimuth. In addition, spiral density waves excited around the vortex can transport angular momentum, with $\alpha\sim10^{-4}-10^{-2}$ \citep{li+01}.

\begin{figure}
    \centering
    \includegraphics[width=0.5\textwidth]{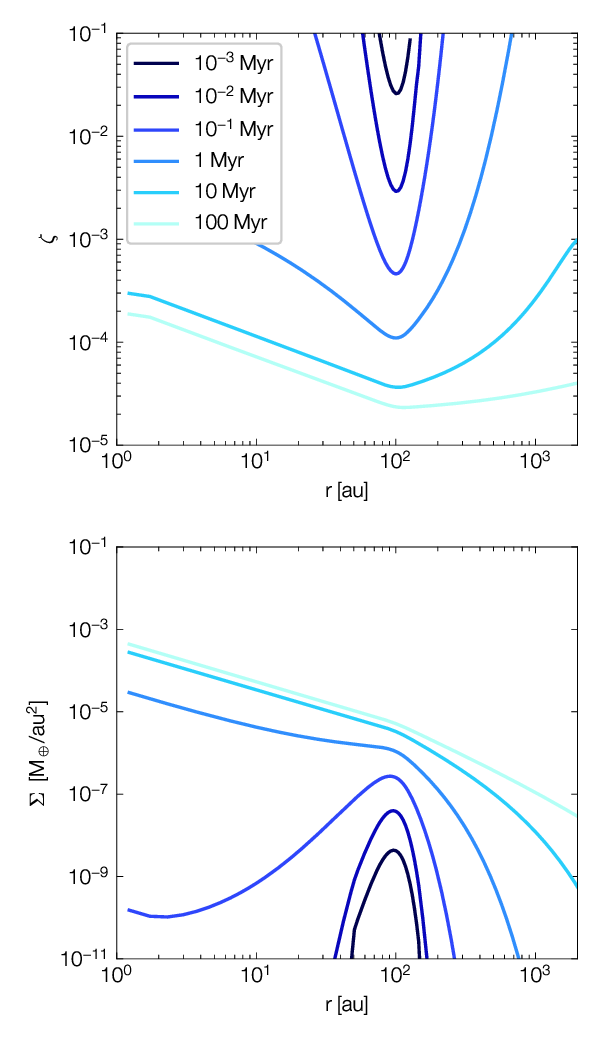}
    \caption{Top panel: vortensity $\zeta$ as a function of radius at different times as evolved forward in the 1D model of \citet{marino+20}. Bottom panel: gas surface density as a function of radius at different times.
    }
    \label{fig:rwi}    
\end{figure}

Figure \ref{fig:rwi} shows the vortensity at different simulation times, using data taken from 1D radial modeling of gas in \citet{marino+22} (their \S4.5), where they consider CO gas released at a rate of 0.1 M$_\oplus$/Myr in a belt centred at 100 AU and 50 AU wide (FWHM), with radial $\alpha = 10^{-2}$. The viscous time scale is 1.5 Myr. It can be seen that all the $\zeta$ profiles possess a local minimum at $r=100$ AU, even for times much longer than the viscous time, and are hence potentially RWI-unstable. 
Note, however, that the RWI condition quoted above is often insufficiently stringent \citep{chang+23}, especially if there is a radial entropy gradient or the density bump is very wide and viscosity is present. It is likely that the very broad vortensity bumps at late times are stable. Direct stability calculations are needed to truly decide on the prevalence of the RWI in debris disks. 

\begin{figure}
    \centering
    \includegraphics[width=0.5\textwidth]{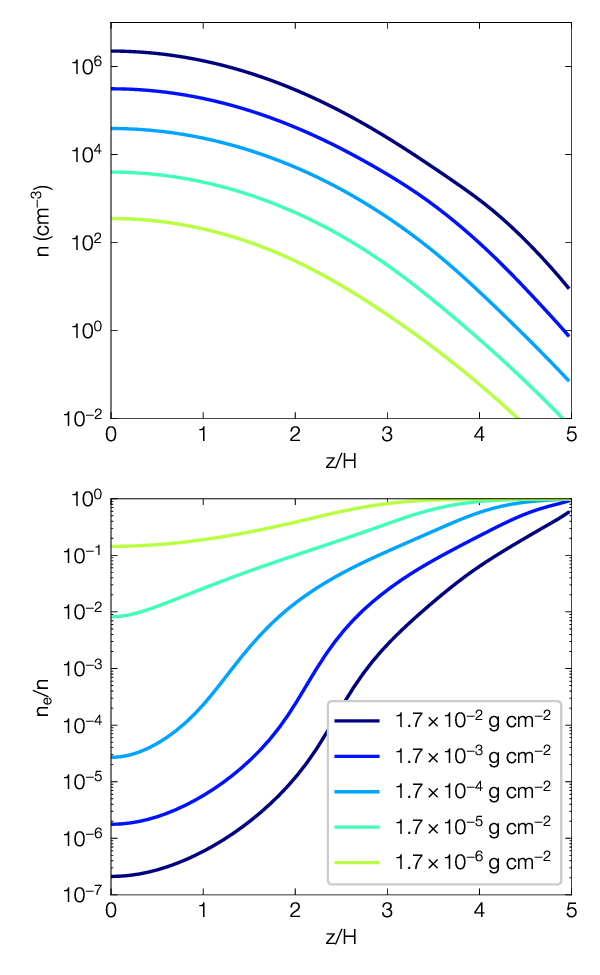}
    \caption{Top panel: neutral number density as a function of height. Bottom panel: ionization fraction as a function of height, calculated by the ratio of electron number density to neutral number density. As earlier, different colours represent different surface densities.}
    \label{fig:seba}    
\end{figure}


\section{Magnetohydrodynamics}\label{sec:MHD}

In this section, we take our fiducial disk models and calculate the ionization fractions and plasma $\beta$ in \S\ref{sec:ip}, as well as the strength of non-ideal MHD effects in \S\ref{sec:nonideal}. MHD mechanisms that can transport angular momentum include magnetic turbulence (MRI) and laminar magnetic stresses, the latter usually associated with a magnetized disk wind. We discuss their application to debris disk gas in \S\ref{sec:amt_mhd}.

\subsection{Ionization and plasma $\beta$}\label{sec:ip}

\subsubsection{Ionization fraction}\label{sec:if}

Utilizing the model described in \S\ref{sec:marino22}, Figure \ref{fig:seba} (top) shows the gas density profile, and Figure \ref{fig:seba} (bottom) shows the ionization fraction $x_\mathrm{e}=n_\mathrm{e}/n$ as functions of vertical height. At low surface densities, the ionization is high even at the midplane, with $x_\mathrm{e}\sim 10^{-2}-10^{-1}$. For high surface densities, the ionization is low, with $x_\mathrm{e}\sim 10^{-7}-10^{-5}$ at the midplane; it is expected that non-ideal MHD effects will set in in this environment.
The maximum ionization is achieved towards the disk surface, which is $x_\mathrm{e}=0.5$ because carbon atoms are entirely ionized but oxygen atoms remain neutral, due to the lower first ionization potential of carbon.

\subsubsection{Plasma $\beta$}\label{sec:beta}

\begin{figure*}
    \centering
    \includegraphics[width=1\textwidth]{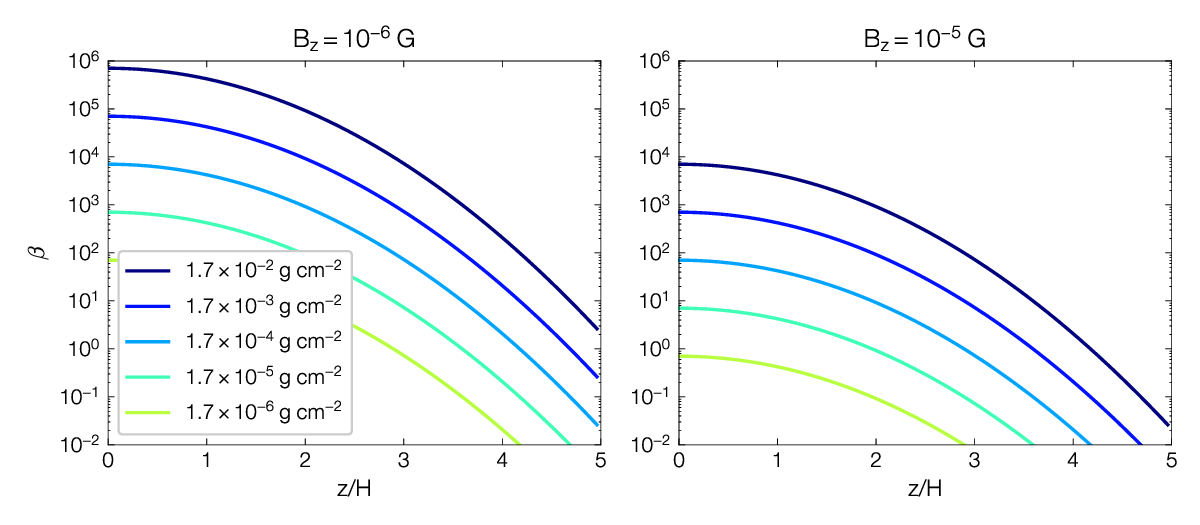}
    \caption{The plasma $\beta$ parameter as a function of height. Left: $B_z=10^{-6}$ G. Right: $B_z=10^{-5}$ G.}
    \label{fig:beta}    
\end{figure*}

The disk magnetization is parametrized by the plasma $\beta$, which is defined as the ratio of the gas pressure to the magnetic pressure, 
\begin{equation}
\beta=\frac{P}{P_B}=\frac{\rho\mathbb{R}T/\mu}{B_z^2/8\pi},
\label{eq:beta}
\end{equation}
where $\mu=14$ is the mean molecular weight (equipartition between carbon and oxygen).
Figure \ref{fig:beta} shows plasma $\beta$ as a function of height for different gas surface densities. The left panel is with a magnetic field strength of $B_z=10^{-6}$ G and right panel of $B_z=10^{-5}$ G. As the magnetic field strength in debris disks is unconstrained (see discussion in \S\ref{sec:origin}), these values are not necessarily representative, though they do correspond to typical solar interplanetary values. Note that the excitation of linear MRI modes require $\beta\gtrsim 1$ in ideal MHD \citep{lk22}.

\subsection{Non-ideal MHD effects}\label{sec:nonideal}

The ideal MHD regime holds only when gas is fully ionized and the gas is frozen into the magnetic field. Non-ideal MHD physics arises when the gas is partially ionized, which is the case in debris disks (Figure \ref{fig:seba}; bottom). The non-ideal MHD effects are manifested in the generalized Ohm's Law. The current density $\bj$ associated with the electric field $\bE$ writes \citep{parks91}
\begin{equation}
\bj = \bm{\sigma}\cdot \bE,
\end{equation}
where the conductivity tensor is 
\begin{equation}
\bm{\sigma} = 
\begin{pmatrix}
\sigma_\mathrm{P} & \sigma_\mathrm{H} & 0 \\
-\sigma_\mathrm{H} & \sigma_\mathrm{P}  & 0 \\ 
0 & 0 & \sigma_\mathrm{O} 
\end{pmatrix}
.
\end{equation}
The Ohmic, Hall,and Pederson conductivities are denoted by $\sigma_\mathrm{O}$, $\sigma_\mathrm{H}$, and $\sigma_\mathrm{P}$, respectively. 
Then, the non-ideal diffusivities for Ohmic resistivity, Hall drift, and ambipolar diffusion are written as 
\begin{equation}
\eta_\mathrm{O} = \frac{c^2}{4\pi} \bigg[\frac{1}{\sigma_\mathrm{O}} \bigg], \quad \eta_\mathrm{H} = \frac{c^2}{4\pi} \bigg[\frac{\sigma_\mathrm{H}}{\sigma^2_\mathrm{H} + \sigma^2_\mathrm{P} } \bigg],
\end{equation}
\begin{equation}
\eta_\mathrm{A} = \frac{c^2}{4\pi} \bigg[\frac{\sigma_\mathrm{P}}{\sigma^2_\mathrm{H} + \sigma^2_\mathrm{P} } \bigg] - \eta_\mathrm{O}.
\end{equation}

For a three-component fluid composed of electrons, ions, and neutrals, the diffusitivies are expressed as
\citep{maeda77,bai17}
\begin{equation}
\eta_\mathrm{O} = \frac{m_\mathrm{e} c^2}{4\pi \mathrm{e}^2} \langle\sigma v\rangle_\mathrm{en} x_\mathrm{e}^{-1},
\end{equation}
\begin{equation}
\eta_\mathrm{H} = \frac{cB}{4\pi \mathrm{e}n}x_\mathrm{e}^{-1},
\end{equation}
\begin{equation}
\eta_\mathrm{A} = \frac{B^2}{4\pi\langle\sigma v\rangle_\mathrm{in} m_\mathrm{n}n^2}x_\mathrm{e}^{-1},
\end{equation} 
where $m_\mathrm{e}$ and $m_\mathrm{n}$ are the molecular masses of electrons and neutrals, $\mathrm{e}$ is elementary charge, $x_\mathrm{e}$ the ionization fraction, and $\langle\sigma v\rangle_\mathrm{en}$ and $\langle\sigma v\rangle_\mathrm{in}$ are the momentum exchange rate coefficients in electron-neutral and ion-neutral collisions, respectively, where angle brackets represent an average over a Maxwellian distribution function for the relative velocity $v$.

\subsubsection{Ohmic resistivity}

Ohmic resistivity issues from electron-neutral collisions. It can erase magnetic fluctuations on length scales shorter than $\sim\eta_\mathrm{O}/\VA$, where $\VA=B/(4\pi\rho)^{1/2}$ is the Alfv\'{e}n speed. To inspect the operation of the MRI with resistivity, we define the Ohmic magnetic Reynolds number
\begin{equation}
\mathrm{Rm} \equiv \frac{c_s\mathrm{H}}{\eta_\mathrm{O}}.
\end{equation} 
This quantity is independent of the magnetic field strength and has a minimum $\mathrm{Rm}\sim10^8$ at the midplane for the largest surface density model. Local linear analysis finds the condition for the onset of MRI in resistive disks to be $1\lesssim\beta\lesssim\mathrm{Rm}^2$ \citep{lk22}. In Figure \ref{fig:beta}, $B_z=10^{-6}$ G gives rise to a maximum beta at the midplane of $\beta\sim10^{6}\ll\mathrm{Rm}^2\sim10^{16}$. Thus, Ohmic resistivity is likely to play a negligible role in debris disks.

\subsubsection{Hall drift}\label{sec:hall}

\begin{figure}
    \centering
    \includegraphics[width=0.5\textwidth]{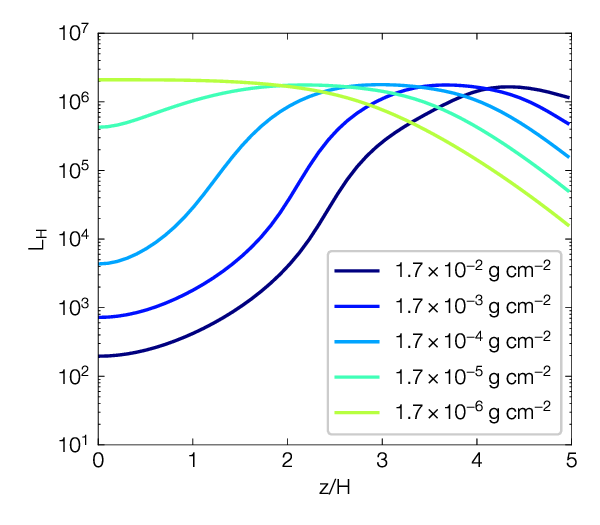}
    \caption{Hall Lundqvist number as a function of height for different surface densities (colours).}
    \label{fig:hall}    
\end{figure}

The ion-electron drift is known as the Hall effect. 
The Hall Els\"{a}sser number is defined as 
\begin{align}
\mathrm{Ha} &\equiv \frac{\VA^2}{\eta_\mathrm{H}\Omega} .
\end{align}
The Hall Lundqvist number, independent of the magnetic field strength, is defined as 
\begin{equation}
L_\mathrm{H} \equiv \frac{\VA H}{\eta_\mathrm{H}}.
\end{equation} 
The effect of Hall drift on disk dynamics depends on the magnetic polarity that co- or counter-aligns with the angular momentum of the disk. However, when $\mathrm{|Ha|}>1$, which is the case for $B_z=10^{-6}$ G and above, the MRI instability criterion reduces to a slightly modified version of its standard form, $\beta\lesssim 1+\mathrm{Ha}^{-1}$. This can be reframed as 
\begin{equation}
\beta\lesssim\frac{1}{4}\big(\sqrt{2}L_\mathrm{H}^{-1}+\sqrt{2L_\mathrm{H}^{-2}+4} \big)^2 
\end{equation}\citep[see Eq.~(49) in][]{lk22}.
Figure \ref{fig:hall} demonstrates that the Lundqvist number $L_\mathrm{H}\gg 1$ for all our disk models. As a result, the criterion reduces to the ideal MHD limit, where the MRI linear modes are only suppressed when $\beta\lesssim1$. Observing $\beta$ in Figure \ref{fig:beta}, the MRI will be stabilized towards the disk surface. 



\subsubsection{Ambipolar diffusion}\label{sec:ad}

\begin{figure*}
    \centering
    \includegraphics[width=1\textwidth]{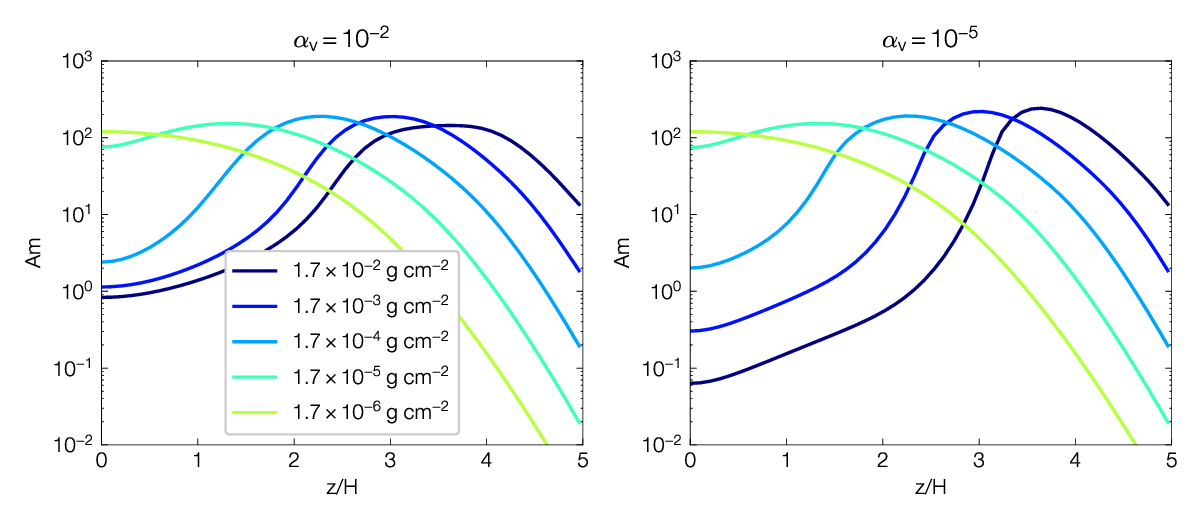}
    \caption{Ambipolar Els\"{a}sser number as a function of height at different gas surface densities (colours). Left: strong vertical diffusion $\alpha_v=10^{-2}$. Right: weak vertical diffusion $\alpha_v=10^{-5}$.}
    \label{fig:am}    
\end{figure*}

Charged particles are coupled to magnetic fields and move relative to the neutrals, but suffer from friction due to ion-neutral collisions. Ambipolar diffusion arises from the drift between ions and neutrals. The Ambipolar Els\"{a}sser number is defined by
\begin{align}
{\rm Am} \equiv\frac{\VA^2}{\eta_\mathrm{A}\Omega}.
\end{align}
Assuming $m_i\approx m_n$, the Els\"{a}sser number can be written as, 
\begin{align}\label{eq:am}
{\rm Am}&\approx\frac{\gamma_i\rho_i}{\Omega} \\ \nn
&\approx 4.1\times10^2x_e\ \Bigg[\frac{\langle\sigma v\rangle_\mathrm{in}}{10^{-9}\ \mathrm{cm^3\ s^{-1}}}\Bigg] \Bigg[\frac{n}{10^2\ \mathrm{cm}^{-3}}\Bigg],
\end{align}
 where $\rho_i$, $m_\mathrm{i}$ are ion density and ion molecular mass, respectively, and $\gamma_i=\langle\sigma v\rangle_\mathrm{ni}/(m_\mathrm{i}+m_\mathrm{n})$ is the drag coefficient of neutral-ion collisions. The momentum exchange rate coefficients $\langle\sigma v\rangle_\mathrm{ni}$ between C$^+$ and CO, CI, OI can be found in Appendix \ref{app:rc}. Eq. \eqref{eq:am} is interpreted as the number of collisions between neutrals and ions per dynamical time. Thereby, when Am is smaller (greater) than unity, the ambipolar diffusion is strong (weak). Figure \ref{fig:am} plots Am versus $z$ for different surface densities and for two different vertical diffusion coefficients. Observing that Am can take significantly low values, we conclude that ambipolar diffusion is the key non-ideal effect in debris disk gas, and one of the main impediments to onset of the MRI. We leave a detailed discussion of its effect to the next section, \S\ref{sec:amt_mhd}.

\subsection{MHD mechanisms of angular momentum transport}\label{sec:amt_mhd}

MHD mechanisms that can transport angular momentum include MRI-induced turbulence, magnetized disk winds, and laminar magnetic stress. Here, we discuss their application to the gas disk based on the calculations carried out in the previous subsections. 

\subsubsection{MRI turbulence}\label{sec:mri}

The MRI, if present, can drive vigorous turbulence and contribute significantly to the angular momentum transport. The MRI operates best in the ideal MHD limit, where the gas is fully ionized, but is weakened by non-ideal MHD effects. In the previous subsection, it was demonstrated that Ohmic resistivity and Hall effect are too weak to become dynamically important. Hence, whether the gas disk is MRI-active depends on the strength of ambipolar diffusion and the plasma $\beta$. While local linear theory indicates the onset of the MRI is unimpeded when $\mathrm{Am}>1$ \citep[e.g.,][]{lk22}, nonlinear simulations suggest that fully developed turbulence only appears when $\mathrm{Am}\gtrsim 10$ and $\beta>1$ \citep{bai11AD}. 

Figure \ref{fig:am} shows ambipolar Els\"{a}sser number as a function of height. We find that the midplane gas in the various disk models falls into one or the other of two distinct scenarios, depending on the surface densities. For high surface densities, the small $\mathrm{Am}\sim 0.1-10$ at the midplane can extinguish or at least weaken the MRI, limiting it to higher disk regions $z>2-3H$ where $\mathrm{Am}\sim 100$ and fortunately the plasma $\beta$ is generically above unity. Note, however, that these regions are collisionless and MRI stability will be altered. Note also that the influence of strong vertical diffusion in aiding the ionization fraction at the midplane \citep{marino+22}. For low surface densities, the ambipolar Els\"{a}sser number is large at the midplane $\mathrm{Am}\sim 100$, and thus MRI turbulence is likely, though one still needs to be sure that the plasma $\beta$ is above unity.

Importantly, the existence or not of the MRI turbulence will feedback strongly on the input parameters of the disk model, such as vertical diffusion and the radial alpha parameter. This feedback can work to further enable the MRI if it is already present, or disable it, if it is not. For example, if MRI turbulence has begun, vertical diffusion will be larger, thus helping to keep the ionisation levels higher (and thus MRI-favourable) at the midplane; meanwhile, radial diffusion of mass will be more efficient, leading to a more dilute disk, and generally higher ionisation levels again. On the other hand, if the disk starts off `dead' with low radial and vertical diffusion, ionisation may remain below the MRI threshold and the disk stays `dead'. This dichotomous behaviour, for the same gas density, might be reflected in hysteresis-type behaviour, though we leave this possibility to future work. 

\subsubsection{Laminar stresses and magnetized disk winds}\label{sec:sw}

Numerical work in the protoplanetary disk context shows that under some circumstances a non-trivial \emph{laminar} MHD state is possible, with strong $R\phi$ magnetic stresses \citep[e.g.,][]{lesur_etal14,bai14,bethune_etal17}. Interestingly, this accreting state can exist independently of a wind. However, it relies on a significant Hall effect, which we can exclude from debris disk gas (\S5.2.2). We hence put this transport route to one side. 

On the other hand, protoplanetary disks are known to transport angular momentum through magnetized disk winds \citep[][]{bs13}, mainly via their accompanying laminar $z\phi$ magnetic stresses. The operation of magnetized disk winds closely depends on the configuration of the background magnetic field, and can work in the presence of ambipolar diffusion \citep[e.g.,][]{gressel_etal15,cb21}. Large-scale, ordered poloidal magnetic fields are critical to launch magnetized disk winds. At the disk surface, a wind is launched and propagates along the magnetic field line, either by a magneto-centrifugal force or by a magnetic pressure gradient \citep{bp82,lyndenb96}. Meanwhile, large-scale correlated azimuthal and poloidal fields exert a torque, by their $z\phi$ Maxwell stress, on the disk surface, which efficiently extracts angular momentum, bestowing it on the gas in the wind. 

Determining the background magnetic configuration and strength in debris disks is challenging because these poperties are closely linked to the largely uncertain origin of the magnetic field (see discussion in \S\ref{sec:origin}). As we argue later, it is unlikely there is any large-scale poloidal field threading debris disks, and thus this transport route is probably absent.


\section{Discussion}\label{sec:disc}

\subsection{Applications to real disk systems}\label{sec:app}

\subsubsection{Vertical mixing of gas} 

The theoretical understanding of gas dynamics can shed light on some key issues in debris disk systems. The most fundamental question is the gas origin. It can be either primordial, a leftover from early Class II protoplanetary disks, or be secondary, released from solids in the planetesimal belt. The secondary origin is particularly promising for dilute gas disks, characterized by low CO abundance. To explain CO-rich systems by the secondary origin route, gas vertical mixing needs to be weak; that way CI at the disk surface can effectively shield midplane CO. But if significant vertical mixing is present, the CI shielding mechanism may be rendered ineffective, suggesting instead that CO-rich systems are of primordial origin. 

Vertical mixing can be provided by some of the instabilities discussed in this paper, especially if they give rise to fully developed turbulence. The MRI and VSI are possible candidates. MRI turbulence leads to random motions of gas, and current numerical simulations of the VSI generically show large-scale coherent vertical oscillations that are effective at mixing gas in the vertical direction.  Importantly, we find that it is the dense CO-rich systems that are most likely to be MRI stable (Section 5.3.1) and thus exhibit the weakest vertical diffusion. In addition, such dense systems are unable to support significant molecular mixing (see Section 4.1). This situation is consistent with the well-shielded scenario and the idea of the secondary origin of CO \citep{kral+19}. The disk, however, may still be VSI unstable, but our preliminary conclusion is that this is unlikely for such dense system, though more detailed calculations of the heating and cooling rates need to be performed in order to better determine the onset of VSI.

\subsubsection{Radial spreading of gas} \label{sec:spread}

If the gas evolves viscously, it will spread in radius and, ultimately, accrete onto the central star. Radial gas spreading could contaminate the atmospheres of embedded planets, enhancing their C/O ratios. It is predicted that an Earth’s atmospheric mass of gas is readily accreted on terrestrial planets in tenuous gas disks \citep{kral+20}. 
The best constraints on the radial distribution of gas come from resolved ALMA observations of CO (and its isotopologues) and CI gas. The spatial distribution of these two species has only been well constrained for 49~Ceti and $\beta$~Pic\footnote{There are other systems with resolved observations of CO and CI (e.g. HD32297), but unfortunately only CO or CI has been analyzed or modeled to constrain its spatial distribution.}. For 49~Ceti it was found that CO is tentatively more compact than CI \citep{higuchi+19}. \cite{kral+19} used a model to fit the carbon gas distribution in HD 131835 and found tentative evidence that CI extends further inwards than CO.

There are mixed results when comparing the gas distribution (CO or CI) relative to the mm-sized dust, perhaps the best measure of potential spreading of gas from its place of production. In some systems the gas distribution is more extended \citep[either towards inner or outer radii, e.g. HD~21997, HD~32297,][]{kospal+13, moor+13}.  
For others, the gas distribution appears less extended \citep[e.g. 49~Ceti, HD~138813,][Mo\'or et al. in prep]{hughes+18, higuchi+19}. 
In others, they are consistent with each other \citep[e.g. $\beta$~Pic, HD~181327, HD~121617, HD~131488, HD~131835, ][Pawellek et al. in prep]{matra+17, matra+19betapic, marino+16, moor+17, kral+19}, suggesting the gas is not subject to any angular momentum transport process and is quiescent. 

These somewhat confounding observations perhaps reflect the complicated nature of angular momentum transport (or its lack) in debris disk systems. While observed gas spreading could issue from any of the transport mechanisms discussed in \S\ref{sec:amt}, we emphasise that the existence or not of these mechanisms is contingent on several, poorly constrained, properties of the system: transport could be magnetic or hydrodynamic, it could be efficient or completely absent. Moreover, there is the added complication that systems could support more than one state for the same set of parameters (see discussion of `hysteresis' in 5.3.1). Though dissatisfying, it does highlight the need to self-consistently connect parameters such as alpha and vertical mixing to the ionization and other properties of the gas in future modeling.

\subsubsection{Vortices and dust clumps}\label{sec:clump}

The RWI saturates via the formation of vortices (\S\ref{sec:rwi}), which in the protoplanetary disk context is a well studied route to the accumulation of dust grains. In this subsection we explore if a similar effect may occur in debris discs. 

The critical issue to resolve here is the strength of the dynamical coupling between dust and gas, which is quantified by the dimensionless Stokes number $\mathrm{St}$. When the dust size is less than the gas mean free path, the case for debris disks, the Epstein drag regime is applicable, and $\mathrm{St}=\pi\rho_s a\Omega/(\rho c_s) \sim \pi\rho_s a/(2\Sigma_g)$, where $\rho_s$ and $a$ denote the internal density and radius of the dust grains \citep{weidenschilling77}. 
Analytical and numerical studies indicate that dust grains of $\mathrm{St}\lesssim 1$ are the most susceptible to trapping in a vortex \citep{johansen+04,surville+16}, though effects peculiar to debris disc gas \citep[e.g., radiation pressure forcing grains into eccentric orbits; see ][]{Skaf+23} may complicate the picture. While generally it is thought that St$>1$ in debris disks, we now check how dense a disk must be for the dust to feel the effect of any vortex present. A rough criterion is
\begin{equation}
    \Sigma_g \gtrsim 2\times 10^{-4}\left(\frac{\rho_s}{\text{g cm}^{-3}}\right)\left(\frac{a}{\mu\text{m}}\right) \,\text{g cm}^{-2}.
\end{equation}
Thus only rather dense gas disks can couple effectively to the dust.

It so happens that $\beta$ Pic exhibits a prominent dust clump, in SW at a stellocentric distance of $\sim55$ AU, observed in the mid-infrared via continuum thermal emission \citep{telesco+05} and comprising $\mu$m (but not mm) sized dust \citep{telesco+05,birnstiel+18,matra+19betapic,sierra+20}. Molecular gas clumps have also been detected in CO and CI emission, likely co-located with the dust clump \citep{dent+14,matra+17,Cataldi+18}. Several physical mechanisms have been proposed in the literature to explain this feature, including an origin via a RWI-induced vortex \citep{Skaf+23}. However, given the low gas surface densities in $\beta$ Pic, which are estimated to be $\mathrm{\Sigma_g}\sim10^{-7}-10^{-6}$ g cm$^{-2}$\citep{kral+16}, the Stokes number for $\mu$m grains is just too high. 
It should be emphasised that $\beta$ Pic is encircled by a rather tenuous gas disc; denser gas systems, subject to the RWI, may be more capable of interacting and imprinting clumpy structure on the dust component.

\subsection{Origin of the background magnetic field}\label{sec:origin}

The presence, strength and configuration of any background large-scale magnetic fields are critical for the dynamics of the gas, if it is sufficiently well ionised (see \S\ref{sec:amt} and \S\ref{sec:MHD}). To begin, the MRI's onset, saturation, and angular momentum transport are linked to the existence or not of large-scale net magnetic flux threading the gaseous disc \citep[e.g.,][]{zhu18,jacquemin21}. On the other hand, large-scale laminar fields can lead to quasi-steady states that drive significant angular momentum transport, usually alongside an MHD wind, which are favoured if non-ideal effects kill off the MRI \citep[e.g.,][]{bs13,lesur_etal14,gressel_etal15,bethune_etal17,cb21}. Presently, we are only just beginning to constrain debris disc magnetism \citep{hull+22}, and so there are no observational constraints on its magnetic field properties. In this subsection, we instead discuss plausible origin scenarios for a background field and what its properties might be.

Firstly, we consider the interplanetary magnetic field permeating a debris disk system. In the solar system, the Cassini and Voyager 1 and 2 reported an interplanetary magnetic field of $\sim 1-10$ nT, equivalent to $1-10$ $\mu$G, \citep[e.g.,][]{jackman+05,Burlaga+22,tsurutani+22}, which is frozen into the solar wind. While such interplanetary magnetic fields exhibit a large-scale structure \citep[see][]{owens13}, debris disk gas will not feel their effects unless they are well penetrated by the stellar wind plasma. This might be possible in dilute systems, but then the disc gas is likely to be simply blown away in a `belt wind' arising from the interaction \citep{kral+23}. In denser systems it is unclear how well the stellar wind plasma can mix with the debris disc gas; while this may be facilitated by instabilities and turbulence at the disc's bow shock, it seems unlikely that the wind's large-scale magnetic structure can be imprinted on the disc in any coherent way, and certainly not in the form demanded by current protoplanetary disc wind modeling (see previous paragraph). That said, the stellar wind should be able to provide a fund of smaller-scale magnetic fluctuations that might serve as the seeds for local MHD activity.  

 Another possibility is that a background magnetic field might be inherited from the preceding protoplanetary disk stage, with the field `carried over' to the later debris stage by residual ionised gas, if it exists, or any charged grains that evade coagulation into pebbles and planetesimals \citep{umebayashi80,ciolek+93}. The viability of this process, however, is yet to be demonstrated. In any case, the final stages of a protoplanetary disc might witness the effective evacuation of most or even all of its large-scale magnetic flux via ambipolar-assisted disc winds (see e.g. \citet{lesur21}). One is hence tempted to conclude that debris disks will not possess a coherent large-scale magnetic field of any significance via this route. 

 In summary, it is likely that any magnetic field within debris disk gas is internally generated from a small seed, via a dynamo process of some kind, most likely via a form of the MRI. We assume that this dynamo is unable to construct a large-scale magnetic field capable of producing meaningful wind accretion, though this may need to be checked (G. Lesur, private communication). On the other hand, we expect an internally generated field to provide further protection from penetration/erosion by the stellar wind, an effect that may work against the launching of belt winds \citep{kral+23}.

\subsection{Ionization sources}\label{sec:is}

In this subsection, we flag a caveat arising from the calculation of the ionization fractions in our debris disk gas model. Previous work on gas evolution includes only the photoionization of gas by an external stellar and interstellar UV radiation field \citep[e.g.,][]{kral+16,marino+22}.
We note that other processes may be important, e.g. the photoelectric effect of dust grains, cosmic rays, and stellar winds, and give them a brief discussion below.

Dust grains are an essential player in the ionization of protoplanetary disks; free electrons are produced by photoionization and are absorbed onto grains' surfaces, leading to negatively charged dust \citep{umebayashi80}. Tiny submicron-sized grains are more efficient in capturing electrons because of their larger total surface density. 
However, in debris disks the smallest grains ($\sim0.1-10\mu$m, depending on the stellar mass) are blown out due to radiation pressure \citep{krivov+06}, thus significantly diminishing the capture of electrons by this route.
Moreover, in debris disks, electrons are more prone to recombine onto ionized carbon atoms rather than to be absorbed by grains when the gas ionization fraction is relatively high, as tested by Cloudy simulations of $\beta$ Pic \citep{kral+16}.  

Photoelectric charging could also play a role as the dust-to-gas density ratio is greater than unity in debris disks. It is found that photoelectric charging, when balanced by the thermal electron collection current, results in positively charged grains \citep[e.g.,][]{Fernandez+06,bw07,zagorovsky+10}. 
For massive gas disks, we expect that both photoelectric and photoionization processes will weaken, because they both depend on the same UV penetrating photons. 
A detailed study of ionization, including dust, in debris disks shall be conducted in the future, on a case-by-case basis, as it strongly depends on the gas versus dust abundances and radiation sources.

Direct and indirect cosmic ray ionization of carbon can be important for denser disks, in which UV photons are less able to penetrate. We present a detailed discussion in Appendix \ref{app:cri}. Its main conclusion is that, unless there are significant levels of primordial H$_2$, cosmic rays increase the Ambipolar Elsasser number by a factor 5 at most, and then only in the densest disks and with efficient vertical mixing.
Lastly, high-velocity stellar wind protons can turn a small fraction of CO to CO$^+$ \citep{kral+21,kral+23}.

\subsection{Future MHD simulations}\label{sec:}

Global numerical simulations should be performed in the context of debris disks in the future, in order to understand the contribution of MRI on the angular momentum transport, turbulence, radial spreading and vertical mixing of gas. The initial magnetic field configuration is key to understanding the gas dynamics, especially when it comes to determine the launching of magnetized disk winds. Initial configuration of large scale open magnetic fields ensuers winds due to flux freezing at and above the disk surface, while poloidal magnetic loops would hinder the launching of winds. In the context of protoplanetary disks, global simulations incorporating non-ideal MHD effects and magnetized disk winds have been conducted, employing large-scale net poloidal magnetic fields \citep[e.g.,][]{bai17,bethune_etal17,wang_etal19,gressel_etal20,cb21}, as inherited from the primordial molecular cloud \citep[e.g.,][]{galli93,girart+06,girart+09}. 
Future simulations for debris disks could test distinct initial magnetic field setups, since it is demanding to constrain the magnetic field strength and configuration. Future observational constraints on the gas distribution and kinematics would be helpful to probe the magnetic field properties of debris disks.

\section{Conclusion}

We review five mechanisms that can transport angular momentum in the context of debris disk gas: molecular viscosity, hydrodynamic turbulence, magnetohydrodynamic turbulence, magnetized disk winds, and laminar magnetic stress. 
Disk models from \citet{marino22} are employed to describe the spatial distribution of the gas species, including CO, CI, CII, and OI (\S\ref{sec:marino22}), and thus provide us with quantiative estimates.
We summarize the main findings as follows. 

\begin{itemize}

\item The molecular viscosity sets a non-negligible lower limit for angular momentum transport in debris-disk gas. The molecular driven, vertically averaged $\alpha$ ranges from $\sim 10^{-5}$ to $\sim 0.1$ from high to low gas surface densities (\S\ref{sec:mw}). In addition, molecular diffusion may contribute significantly to vertical mixing in more dilute systems, as we expect the kinetic $\alpha_v$ to be $\sim\alpha$.

\item The vertical shear instability is unlikely to drive hydrodynamic turbulence. Only in especially dilute gas regions is the thermal relaxation time sufficiently rapid, but the gas is reasonably well ionised and the MRI and/or magnetic tension will intervene in the VSI's development (\S\ref{sec:vsi}). 

\item In the secondary origin scenario, the gas disk naturally exhibits a bump structure in the surface density, at least early in its evolution. This may generate gas vortices via the Rossby wave instability, but a sufficiently large gas surface density is needed to ensure $\mathrm{St}\lesssim 1$ and hence the formation of dust clumps (\S\ref{sec:rwi} and \S\ref{sec:clump}). 

\item The magneto-rotational instability is controlled by the ambipolar diffusion strength, while Ohmic resistivity and Hall effect are too weak to make a significant impact in our disk model. More dilute gas disks should be MRI turbulent at all heights, while in denser disks the MRI may be confined to the better ionised upper layers $z>2-3H$, though these conclusions rely on a background magnetic field $B_z=10^{-6}-10^{-5}$~G, which is poorly constrained (\S\ref{sec:mri}).

\item Laminar $R\phi$ magnetic stresses can transport angular momentum but rely on a significant Hall effect, and are hence ruled out in the debris disk context. Laminar $z\phi$ magnetic stresses rely on a magnetized disk winds, which in turn requires a large-scale vertical magnetic field (\S\ref{sec:sw}) threading the disk. We are skeptical that such a B-field configuration is possible in debris disks, though future investigation is necessary to decide on this point (\S\ref{sec:origin}). 

\item Of the angular momentum transport mechanisms reviewed, MRI turbulence appears the most robust, but it is only effective in more dilute, and hence better ionised, disks. In denser systems it is likely there is little transport at all. This suggests gaseous debris disks are dichotomous. If a disk begins dilute and MRI turbulent then, via effective vertical mixing and radial spreading, it will remain dilute, well-ionised, and thus MRI-turbulent. If a disk begins dense, poorly ionised, and quiescent, it will remain so; in fact, mass may continue to build up, as it will be inefficiently spread radially (\S\ref{sec:mri}). This dependence on the disk's dynamical history (and on its poorly constrained physical parameters) perhaps explains the extant, somewhat confounding, results on gas spreading (\S\ref{sec:spread}).

\end{itemize}


\section*{Acknowledgements}

We thank Daniele Galli for pointing out where we could obtain the ionisation rate of CI by cosmic rays. We also thank Mark Wyatt and Ya-Ping Li for useful discussions at the earlier stage of this project. CC and HNL acknowledge funding from STFC grant ST/T00049X/1. SM is supported by a Royal Society University Research Fellowship (URF-R1-221669). 

\section*{Data Availability}

The data underlying this article will be shared on reasonable request to the corresponding author.


\appendix
\section{Momentum exchange rate coefficients}\label{app:rc}

\begin{figure}
    \centering
    \includegraphics[width=0.5\textwidth]{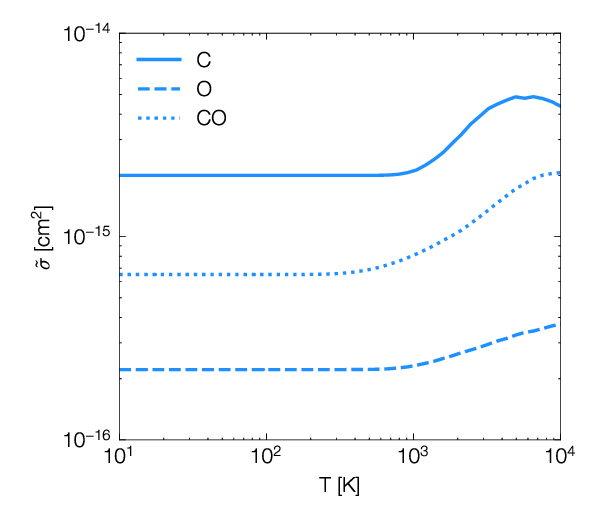}
    \caption{The integration part of eq. \eqref{eq:A1} for C, O, and CO.}
    \label{fig:rate}    
\end{figure}

The momentum exchange rate coefficients of electron-neutral and ion-neutral collisions are calculated here in order to compute Ohmic and ambipolar Els\"{a}sser number. 

Electron–neutral scattering limits the electrical conductivity in low ionization regions. The two-body collision rate coefficient is \citep{draine11}, 
\begin{equation}
\langle\sigma v\rangle = \Bigg[\frac{8k_BT}{\pi\mu}\Bigg]^{1/2} \int_0^\infty \sigma(E) \frac{E}{k_BT} e^{-E/k_BT}\ \frac{dE}{k_BT}  \ ,
\label{eq:A1}
\end{equation} 
where $k_B$ is the Boltzmann constant, $\mu$ the reduced mass of the two collision species, $T$ the thermal euqilibrium temperature, $E$ the center-of-mass energy, and $\sigma$ the reaction cross section. The momentum exchange cross sections $\sigma_\mathrm{mt}(E)$ for e$^--$O, and e$^--$CO collisions are tabulated in \citet{Thomas75,Itikawa90,Itikawa15}. The elastic scattering cross section for e$^--$C is read in \citet{Ronald+69}.
Note that the cross sections measured in laboratory experiments is usually given as a function of energy in the laboratory frame. For electron-neutral collisions, $E \approx E_\mathrm{lab}$. 

In Figure \ref{fig:rate}, the integration part of eq. \eqref{eq:A1} $\tilde{\sigma}=\int_0^\infty \sigma(E) \frac{E}{k_BT} e^{-E/k_BT} \frac{dE}{k_BT}$ is shown for C, O, and CO. When $10 \lesssim T\lesssim 1000$ K, $\tilde{\sigma}_\mathrm{e^-,C} = 2\times10^{-15}\ \mathrm{cm}^2$,
$\tilde{\sigma}_\mathrm{e^-,O} = 2\times10^{-16}\ \mathrm{cm}^2$,
$\tilde{\sigma}_\mathrm{e^-,CO} = 6.5\times10^{-16}\ \mathrm{cm}^2$.
The corresponding rate coefficients can then be computed by 
\begin{equation}
\langle\sigma v\rangle_\mathrm{e^-} = 4.14\times10^5 \ T^{1/2}\
\tilde{\sigma} \ \mathrm{cm^3 \ s^{-1}}\ ,
\end{equation}  
and at $T=50$ K,
\begin{equation}
\langle\sigma v\rangle_\mathrm{e^-,C} =  5.85\times10^{-9}\ \mathrm{cm^3 \ s^{-1}}\ ,
\end{equation} 
\begin{equation}
\langle\sigma v\rangle_\mathrm{e^-,O} =  5.85\times10^{-10}\ \mathrm{cm^3 \ s^{-1}}\ ,
\end{equation} 
\begin{equation}
\langle\sigma v\rangle_\mathrm{e^-,CO} = 1.9\times10^{-9} \ \mathrm{cm^3 \ s^{-1}}\ .
\end{equation} 

The ion–neutral collision rates determine the ambipolar diffusion strength. The momentum transfer rate coefficients for ion-neutral collisions between C$^+$ and C, O, CO are calculated by eq. (2.34) and (2.39) of \citet{draine11}, with the tabulated dipole polarizabilities \citep{maroulis96,polar18}. This yields:
\begin{equation}
\langle\sigma v\rangle_\mathrm{C^+,C}=9.57\times10^{-10} \ \mathrm{cm^3 \ s^{-1}},
\end{equation} 
\begin{equation}
\langle\sigma v\rangle_\mathrm{C^+,O}=1.49\times10^{-9} \ \mathrm{cm^3 \ s^{-1}},
\end{equation} 
\begin{equation}
\langle\sigma v\rangle_\mathrm{C^+,CO}=1.35\times10^{-9} \ \mathrm{cm^3 \ s^{-1}}.
\end{equation}

\section{Cosmic ray ionization of carbon}\label{app:cri}

For tenuous debris disks such as $\beta$ Pic, stellar and interstellar UV should dominate the ionization processes. Meanwhile, denser disks may become optically thick to ionising UV photons at some height. In these circumstances, cosmic ray ionization might be significant and may be the leading ionization source towards the midplane \citep{umebayashi_nakano81}. To test this, we focus on the direct and indirect cosmic ray ionization of carbon as it has a lower first ionization potential than oxygen. 

We implemented this new source of ionisation for carbon in exogas\footnote{https://github.com/SebaMarino/exogas} \citep{marino+22}, and evaluate two cases. First, we treat the case where the gas is of secondary origin and thus there is no H$_2$. Here the dominant effect of cosmic rays is via the direct ionisation of carbon at a rate of 1.94 relative to the standard value for H$_2$ \citep[i.e. a CI ionisation rate of $1.94\times10^{-16}$ s$^{-1}$,][]{prasad1980, heays2017}. Second, we treat the case where the gas is of a primordial origin and there are orders of magnitude more H$_2$ than CI. Now the dominant effect of cosmic rays on CI is the indirect ionisation via the excitation of H$_2$ and H followed by the emission of UV photons capable of ionising CI. In this case, we use the tabulated value in \citet{heays2017} of $2.6\times10^{-14}$ s$^{-1}$. Note that this value depends on the abundance of dust and CI and it assumes the disc is optically thick to ionising UV photons (valid in the primordial scenario). While the abundance of dust and CI could be different than in the context studied by \citet[CI abundance of $10^{-5}-10^{-4}$,][]{heays2017}, in the absence of a better estimate we use their value. 

\begin{figure*}
    \centering
    \includegraphics[width=1\textwidth]{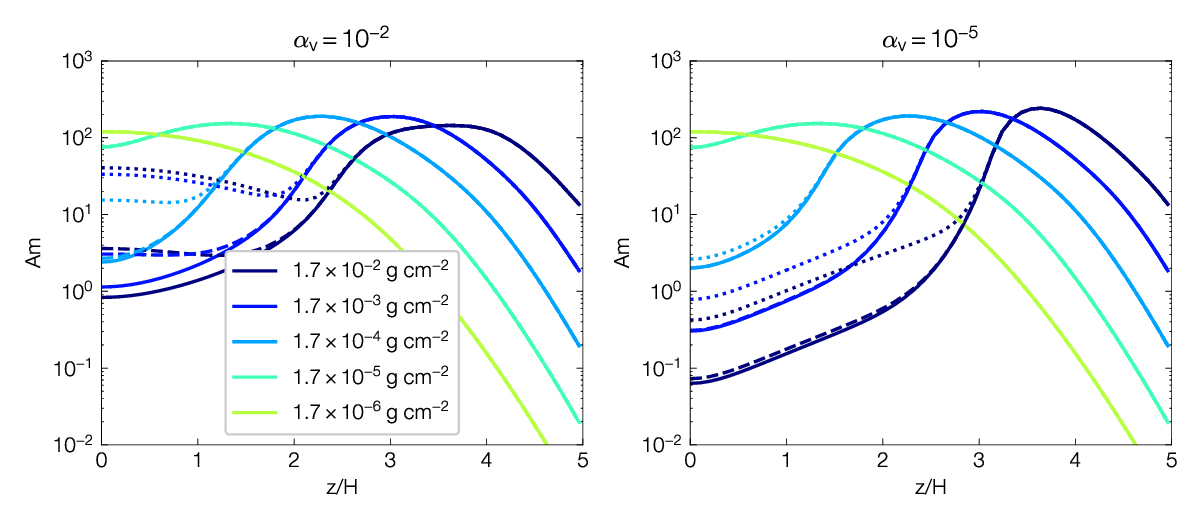}
    \caption{
    Ambipolar Els\"{a}sser number Am for direct and indirect cosmic ray ionization. Solid line: no cosmic ray ionization. Dashed line: direct cosmic ray ionization. Dotted line: indirect cosmic ray ionization of carbon by the excitation of H$_2$ and H. Left: strong vertical diffusion $\alpha_v=10^{-2}$. Right: weak vertical diffusion $\alpha_v=10^{-5}$. Different colours indicate different surface densities.
    }
    \label{fig:cr}    
\end{figure*}

Figure \ref{fig:cr} shows the Ambipolar Els\"{a}sser number Am for direct and indirect cosmic ray ionization.  
For strong vertical mixing ($\alpha_v=10^{-2}$), the impact of cosmic rays is notciceable (left panel). Direct cosmic ray ionization yields a maximum amplification of ionization fraction by a factor of $\sim 5$ for dense disks at the midplane. Indirect cosmic ray ionization (due to the excitation of H$_2$) has a stronger effect, a factor $\sim 30-40$ increase of Am at the midplane. It indicates the entire disk can be MRI turbulent. Of course, indirect ionization requires the presence of large amounts of H$_2$ which would only be the case only if the gas were of primordial origin.

For weak vertical mixing ($\alpha_v=10^{-5}$), the impact of cosmic ray is limited. As mixing is inefficient, the amount of CI in the midplane is low, which lowers the ionisation fraction and Am in the midplane.
For dilute disks, the ionization by UV photons dominates over the cosmic ray, and hence the three curves almost overlap for both weak and strong vertical diffusion. 

\bibliographystyle{mnras}
\bibliography{3d} 

\bsp	
\label{lastpage}
\end{document}